\documentclass[sigconf]{acmart}
\AtBeginDocument{%
  }

\setcopyright{acmlicensed}
\copyrightyear{2026}
\acmYear{2026}
\acmDOI{10.1145/3744916.3773215}
\acmConference[ICSE '26]{IEEE/ACM 48th International Conference on Software Engineering}{April 12--18,
  2026}{Rio De Janeiro, Brazil}
\acmISBN{978-1-4503-XXXX-X/2018/06}

\usepackage{multirow}

\usepackage{url}

\usepackage{afterpage}
\usepackage{graphicx}

\usepackage[colorinlistoftodos]{todonotes}

\usepackage[dvipsnames]{xcolor}

\usepackage{soul}

\newif\ifreview
\reviewfalse 

\ifreview
  \newcommand{\newText}[1]{{\color{blue}#1}}
  \newcommand{\delete}[1]{{\color{red}\st{#1}}}
\else
  \newcommand{\newText}[1]{#1}
  \newcommand{\delete}[1]{}
\fi

\usepackage{listings}
\definecolor{darkyellow}{RGB}{125, 99, 45}
\definecolor{blue}{RGB}{0, 24, 255}
\definecolor{purple}{RGB}{175, 0, 219}
\definecolor{lightblue}{RGB}{38, 128, 153}
\definecolor{green}{RGB}{8, 128, 0}
\definecolor{red}{RGB}{163, 21, 21}
\definecolor{lightgrey}{RGB}{204, 204, 204}
\definecolor{red}{RGB}{163, 21, 21}
\definecolor{errorred}{RGB}{241,156,153}

\lstdefinestyle{Python}{
  language=Python,
  basicstyle=\small\ttfamily,
  basewidth=0.55em,
  numbers=left,
  numberstyle=\small\color{lightgray},
  tabsize=2,
  columns=fixed,
  showstringspaces=false,
  showspaces=false,
  showtabs=false,
  keepspaces,
  commentstyle=\color{green},
  stringstyle=\color{red},
  keywordstyle=\color{purple},
  morekeywords={with,as,then},
  deletekeywords={def,not,in,and,or},
  keywords={[2]@invariant,
  deque, range,
  lasapp,
  z3,
  ProbabilisticProgram, HamiltonianMC,
  Warning,
  Interval,
  Implies, And, Not, Or, satisfiable,
  pymc, pm, Model, Normal,
  pyro, torch, dist, Bernoulli, Beta, Uniform, Gamma, InverseGamma, Categorical, HalfCauchy, bool
  },
  keywordstyle={[2]\color{lightblue}}, 
  keywords={[3]@invariant,
  len, append, popleft, items, keys,
  get_model,
  get_guide,
  get_data_dependencies,
  get_random_variables,
  get_control_dependencies,
  get_control_parents,
  get_call_graph,
  estimate_value_range,
  get_path_condition,
  infer_distribution_properties,
  has_random_control_deps,
  get_dist_constraint,
  check_disjointness,
  Z3Symbol, SExpr,
  is_discrete,
  already_processed,
  mark_processed,
  is_subset,
  is_subset_of,
  check,
  filter,
  sample, observe, Deterministic,
  tensor, ones, abs,
  @bm, random_variable,
  },
  keywordstyle={[3]\color{darkyellow}},
  keywords={[4]@invariant,def,not,and,or,in,True,False},
  keywordstyle={[4]\color{blue}},
}

\usepackage{cleveref}
\usepackage{enumitem}
\usepackage{placeins}

\newcommand{\toolname}{\textsc{InferLog Holmes}}
\begin{document}

\title{Online and Interactive Bayesian Inference Debugging}

\author{Nathanael Nussbaumer}
\authornote{Both authors contributed equally to this research.}
\email{nathanael.nussbaumer@tuwien.ac.at}
\affiliation{%
  \institution{TU Wien}
  \city{Vienna}
  \country{Austria}
}

\author{Markus Böck}
\authornotemark[1]
\email{markus.h.boeck@tuwien.ac.at}
\affiliation{%
  \institution{TU Wien}
  \city{Vienna}
  \country{Austria}
}

\author{Jürgen Cito}
\email{juergen.cito@tuwien.ac.at}
\affiliation{%
  \institution{TU Wien}
  \city{Vienna}
  \country{Austria}
}


\renewcommand{\shortauthors}{Nussbaumer et al.}

\begin{abstract}
 Probabilistic programming is a rapidly developing programming paradigm which enables the formulation of Bayesian models as programs and the automation of posterior inference. It facilitates the development of models and conducting Bayesian inference, which makes these techniques available to practitioners from multiple fields. Nevertheless, probabilistic programming is notoriously difficult as identifying and repairing issues with inference requires a lot of time and deep knowledge. Through this work, we introduce a novel approach to debugging Bayesian inference that reduces time and required knowledge significantly. We discuss several requirements a Bayesian inference debugging framework has to fulfill, and propose a new tool that meets these key requirements directly within the development environment. We evaluate our results in a study with 18 experienced participants and show that our approach to online and interactive debugging of Bayesian inference significantly reduces time and difficulty on inference debugging tasks.
\end{abstract}

\ccsdesc[500]{Software and its engineering~Software testing and debugging}
\ccsdesc[500]{Mathematics of computing~Bayesian computation}
\ccsdesc[500]{Human-centered computing~Interactive systems and tools}

\keywords{
Probabilistic programming, Bayesian inference, debuggging
}

\received{18 July 2025}
\received[accepted]{17 October 2025}

\maketitle

\todototoc

\section{Introduction}

Probabilistic programming is a rapidly developing programming paradigm which enables the formulation of Bayesian models as programs and the automation of posterior inference.
In Bayesian modeling we specify relationships between observable variables and unobservable \emph{latent} variables.
Knowledge about the latent variables before collecting data can be incorporated by defining a \emph{prior distribution} over these variables.
Once data has been collected, we are interested in the distribution over the latent variables \emph{conditioned} on observing the data -- the \emph{posterior distribution}.

By abstracting away much of the inference machinery and letting the user focus on model building, probabilistic programming aims to make Bayesian inference more accessible to practitioners who are not experts in inference algorithms.
However, this separation often makes it difficult to resolve issues in probabilistic programming, because its output -- the approximation to the posterior -- heavily depends on the interplay of model and inference.
Furthermore, even if a probabilistic program runs without raising exceptions and is bug-free in the traditional programming sense, its output may still be poor.
Currently, practitioners require deep knowledge of probabilistic programming systems to resolve issues\newText{. Furthermore, due to the heavy computational demands of Bayesian inference algorithms \cite{bishop2006pattern, ADVI, StochasticVariationalInference, sountsov2024runningmarkovchainmonte},}\delete{and} development cycles are slow, as one has to wait for a potentially long inference time before being able to analyzing the output.

In this work, we address the aforementioned challenges by designing, implementing, and evaluating a development environment for debugging probabilistic programs.
We derive six design requirements for such an environment with the most central ones being \emph{online debugging} and \emph{interactive debugging}.
The former refers to being able to analyze the output of a probabilistic program \emph{during} its execution, while the latter requires a graphical user interface for diagnosing issues.
We realized these requirements in the implementation of \toolname{} - our debugger specialized for workflows using so-called MCMC inference algorithms.
In a user-study we found that practitioners are more effective at detecting and resolving issues with our environment.

In summary, the contributions of this work are
\begin{itemize}
    \item the design of an online and interactive debugger for probabilistic programs,
    \item the open-source implementation of \toolname{} -- a debugger specialized for MCMC workflows (\url{https://github.com/ipa-lab/InferlogHolmes-Appendix}),
    \item a user-study with 18 participants evaluating the debugging experience with \toolname{},
    \item a quantitative and qualitative analysis of the effectiveness of \toolname{}.
\end{itemize}

\section{Debugging in Probabilistic Programming}\label{sec:debugging-insights}

We studied the plethora of literature on diagnosing problems in Bayesian inference \cite{brooks2011handbookmcmc, Hamada2022BugSystems,Gorinova2016ANovices,GabryVisualizationWorkflow,Vats2020AnalyzingOutput,gelman1995bayesiandataanaysis}, as well as the provided tooling of modern probabilistic programming systems \cite{abril2023pymc, carpenter2017stan, bingham2019pyro, ge2018turing}, and summarize our findings in three statements \textit{DI1}-\textit{DI3} termed \emph{debugging insights}.
We illustrate the process of debugging in probabilistic programming and introduce our insights at the hand of an example presented in Rubin~\cite{Rubin1981EstimationExperiments}.

Namely, we consider a model where researchers investigated whether short-term coaching can lead to increasing SAT scores (examinations designed to be resistant against such short-term efforts).
For each school $i$, the researchers have gathered data on the differences in test scores $y_i$ along with a standard error estimate~$\sigma_i$.
To account for differences and similarities of the effect between schools, we model the experiment mathematically as follows:
\begin{align*}
\mu &\sim \textnormal{Normal}(0,10) &\dots& \quad \text{overall treatment effect} \\
\tau & \sim \textnormal{HalfCauchy}(5) &\dots& \quad \text{effect variance between schools} \\
\theta_i &\sim \textnormal{Normal}(\mu,\tau)  &\dots& \quad \text{effect for school }i \\
y_i  &\sim \textnormal{Normal}(\theta_i,\sigma_i)  &\dots& \quad \text{observed effect for school }i
\end{align*}
The appeal of probabilistic programming is that we can translate this mathematical description almost one-to-one to a programmatic formulation, for instance, to PyMC \cite{abril2023pymc}, as below:
\begin{lstlisting}[style=Python, numbers=right, numbersep=0pt, resetmargins=true]
with pm.Model() as eight_schools:
  mu = pm.Normal("mu",mu=0,sigma=10)
  tau = pm.HalfCauchy("tau",beta=5)
  for i in range(8):
    theta = pm.Normal(f"theta{i}",mu=mu,sigma=tau)
    pm.Normal(f"y{i}",mu=theta,sigma=sigma,observed=y)
\end{lstlisting}
Separate from the model definition, we specify our inference routine, where we use the general-purpose HMC algorithm:
\begin{lstlisting}[style=Python, numbers=none, numbersep=0pt, resetmargins=true]
with eight_schools:
  step = pm.HamiltonianMC(step_scale=0.2)
  idata = pm.sample(3000,tune=100,step=step,chains=4)
\end{lstlisting}
The drawback of probabilistic programming systems is the fact that, while the programs look fairly simple, a lot of complex inference machinery is abstracted away and when inference does not work out-of-the-box, then finding the underlying issues is very challenging.
In fact, the presented code would not produce a good approximation of the true posterior.

To improve the result, first, we would need to replace the for loop in lines 4-6 with
\begin{lstlisting}[style=Python, numbers=none, numbersep=0pt, resetmargins=true]
Z = pm.Normal("Z",mu=0,sigma=1,shape=(8,))
theta = pm.Deterministic("theta", mu + tau * Z)
pm.Normal(f"ys",mu=theta,sigma=sigmas,observed=ys)
\end{lstlisting}
This change accelerates inference by batching the school effects~$\theta_i$ and observed data~$y_i$, $\sigma_i$ into vectors.
But more importantly, it reparametrizes the model such that the scale of all variables is independent of other latent variables.
Without going into too much detail, HMC cannot explore the posterior distribution well if the scale of $\theta$ depends on $\tau$ via \texttt{sigma=tau} and it is better to derive theta from a the latent variable \texttt{Z} with fixed scale.

Coming up with this improvement requires deep knowledge about HMC, how it is implemented in the internals of PyMC, and how our model definition affects its behavior.
Such issues are not unique to HMC nor PyMC, and in general, to find these issues we state following debugging insight:
\begin{quote}
\textit{
DI1. Deep understanding of the interplay between inference routine and model required.
}
\end{quote}

The second issue with the presented code lies within the sub-optimal choice of the inference hyper-parameters.
The \texttt{step\_scale} is set slightly too small and in conjunction with a small number of tuning steps, leads to inefficient posterior exploration.
But how would a practitioner identify these issues?

\begin{figure}[h]
    \centering
    \includegraphics[width=0.95\linewidth]{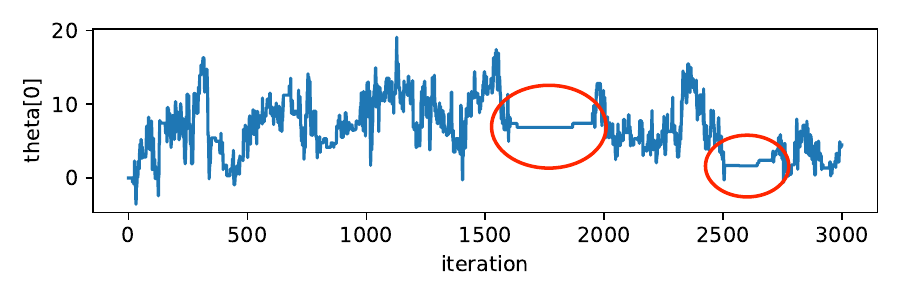}
    \caption{Example of a suspicious trace plot.}
    \label{fig:theta-trace}
\end{figure}

Results produced by Markov-Chain-Monte-Carlo (MCMC) type algorithms like HMC are typically analyzed through means of diagnostic statistics and plots.
For instance, we may find that the so-called trace plot in \Cref{fig:theta-trace}, which shows the value of a latent variable during inference, exhibits suspicious "flat" areas. \newText{This is an indication that the sampling algorithm got "stuck" at some point and that it proposed many values that are not from the true posterior.}
Still, deep knowledge is required to interpret these plots and make changes to the model and inference setup to mitigate the issues.

Furthermore, it is important to highlight that we can perform this analysis only when the inference routine has finished.
As MCMC algorithms are only asymptotically correct, they have to be run for a large number of iterations and sample sizes between tens of thousands and millions are common.
This leads to long inference time and slow model/inference development cycles.
We summarize these insights into following fact:
\begin{quote}
\textit{
DI2. Diagnosing inference results requires deep knowledge of the interplay between modeling and the underlying inference machinery, and is only possible after potentially long inference times.
}
\end{quote}

Lastly, we note that in \Cref{fig:theta-trace}, we conveniently showed a specific diagnostic plot that is indicative of issues in the presented code.
However, there is a large number of diagnostic statistics and plots that should be considered for each latent variable individually.
Typically, the same inference routine is run in parallel several times such that issues may manifest in one run but not in others.
With the configuration \texttt{chains=4}, we run inference four times and our model defines ten latent variables.
If there are $N$ suitable diagnostic plots to consider, the total number of plots to analyze is $40\times N$.
This even neglects the class of diagnostics tools designed to analyze \emph{pairs} of random variables.
With our last debugging insight we summarize that the number of diagnostics statistics and plots to consider grows quickly.
\begin{quote}
\textit{
DI3. The debugging exploration space for probabilistic programs is very large.
}
\end{quote}
The challenges for practitioners discussed in this section are not unique to PyMC, the presented model or chosen inference algorithm.
Instead, they are universally encountered in probabilistic programming and in the next sections, we will introduce our approach to improve the debugging experience.




\section{Debugger Design}
In this section, we derive six \emph{design requirements} \textit{REQ1}-\textit{REQ6} for PPL debuggers derived from the debugging insights \textit{DI1}-\textit{DI3} introduced in \Cref{sec:debugging-insights}.
In this context, debugging refers to locating and fixing issues in the model definition or inference result.
While these requirements are not tailored to any specific PPL nor inference algorithm, we implemented a tool specifically for debugging MCMC inference in PyMC based on them, named \toolname{}.
We describe the implementation of \toolname{} alongside the design requirements  and present the evaluation of \toolname{} in the subsequent sections.


\subsection{REQ1: Multiple-View Integrated Development Environment}
In contrast to a traditional debugger, which allows practitioners to set breakpoints and step through the program line by line, locating and debugging Bayesian inference issues comes down to analyzing the quality of the posterior distribution approximation.


The quality of this approximation is tightly coupled to the source code of a probabilistic program via the model definition, choice of inference method and various hyper-parameters. 
The user is required to understand the interplay of these components for fixing issues as stated in \textit{DI1}.
Having a secondary view for these analysis tools next to the primary source code editor gives the practitioner fast feedback on how a change in the source code is reflected in the inference result.

We have implemented \toolname{} as a VSCode extension which splits the screen in half.
Fig.~\ref{fig:example_workflow} shows that in addition to the source code view on the left, we have a secondary panel on the right for displaying debugging information.
This secondary panel provides three tabs:
1) a \emph{Model View} which visualizes the dependencies between random variables in the model graphically;
2) a \emph{Live Debugging View} which displays intermediate results while inference is running, see \Cref{sec:req-on-line};
3) a \emph{Warnings View} which lists potential issues detected by analyzing the incoming inference results on the fly, see \Cref{sec:req-heuristic}.

\subsection{REQ2: Contextualized Workflows and Knowledge Base}
There is a lot of information about recommended workflows for specific inference approaches throughout literature and educational resources.
As noted in \textit{DI1}, for fixing inference issues it is crucial to understand the interplay of the employed inference algorithm and the implemented model.
Thus, a debugger has to provide analysis tools and workflows in the context of the chosen inference algorithm and supplies the practitioner only with information relevant to their current implementation.

To fulfill \textit{REQ2}, our tool detects the inference algorithm specified in the source code and adapts its visualizations and background analysis accordingly.
For instance, while for MCMC algorithms like HMC trace plots are suitable (see \Cref{fig:theta-trace}), they do not make sense in a \emph{Variational Inference}~\cite{StochasticVariationalInference} context, where the convergence of the so-called ELBO loss is of interest.
Differences may also be more subtle:
The optimal acceptance rate\footnote{\newText{The acceptance rate measures the proportion of proposed sampled that are accepted during inference. Although it is not a strong standalone indicator, high acceptance rates can signal slow exploration of the posterior and waste of compute resources, while low acceptance rates often hint that the algorithm has trouble staying within regions of high posterior probability.}}, which is used as reference in the background analyses, differs among MCMC algorithms. \newText{} \newText{
Via its modular design, \toolname{} allows us to easily add analyses and configure existing analyses for given inference algorithms.
For instance, we may configure our analyses to the theoretically optimal acceptance rates of the MCMC algorithm at hand.
Additionally, this modularity also enables \toolname{} to generalize to other classes of inference algorithms like variational inference, e.g. by adding plots of the ELBO loss.

}

\subsection{REQ3: Online Diagnostic Tools}\label{sec:req-on-line}
Executing a probabilistic program amounts to running an inference routine for thousands of iterations.
As stated in \textit{DI2}, for complex models this takes a long time.
State-of-the-art probabilistic programming systems only allow you to analyze the inference result after the routine has finished, leading to slow development cycles.
An effective debugger enables such analysis \emph{online}, i.e. before, during, and after running the inference routine.

\newText{In general, this can be achieved by hooking into the inference loop of any PPL. Depending on the language design this could be done in an external package or through direct integration.
To collect the inference state after each iteration for PyMC, we implemented a simple class which wraps a trace object and make use of functionality intended for storing traces on disk during inference.
Through batching on both the language and tool sides, we keep the performance overhead from communication and analysis negligible.
This enables efficient, real-time updates of all visualizations and analyses as results become available on consumer-grade hardware.
}
\delete{By hooking into the inference loop of PyMC with a simple wrapper class realized in an importable Python package and collecting the inference state after each iteration, \toolname{} is able to update all visualizations and analyses as results come in.}

In \toolname{}, the \emph{Debugging View} shows the trace plot and histogram of variables up to the current iteration and the \emph{Warnings View} displays the latest available insights from the background analysis.
This way we enable practitioners to spot issues quickly and allow them to abort inference early if necessary.

\subsection{REQ4: Interactive Diagnostic Tools}
As stated in \textit{DI2} and \textit{DI3}, to analyze the inference result, the practitioner needs to know which diagnostic methods are applicable, typically write a lot of boilerplate code to use them, and repeatedly run the code on a selection of the variables.
To make this process less cumbersome, a debugging tool must integrate such diagnostic tools as a core component and allow the user to interactively perform the same analysis without the need to write code.

To this end, with \toolname{} we enable the user to navigate the interface primarily by mouse clicks.
For instance, in the \emph{Debugging View} variables are selectable similar to browser tabs, the particular MCMC chain can be chosen from a drop-down menu, and the user can jump to more detailed plots by the press of a button.
In the \emph{Warnings View}, initially a condensed summary of the warnings is given, each of which can be expanded to a detailed description.
Lastly, we include links to relevant source code sections throughout the interface to strengthen the connection between code and analysis.

\subsection{REQ5: Heuristic-Guided Exploration of the Debugging Space}\label{sec:req-heuristic}

\begin{figure}[h!]
\centering
\includegraphics[width=0.5\textwidth]{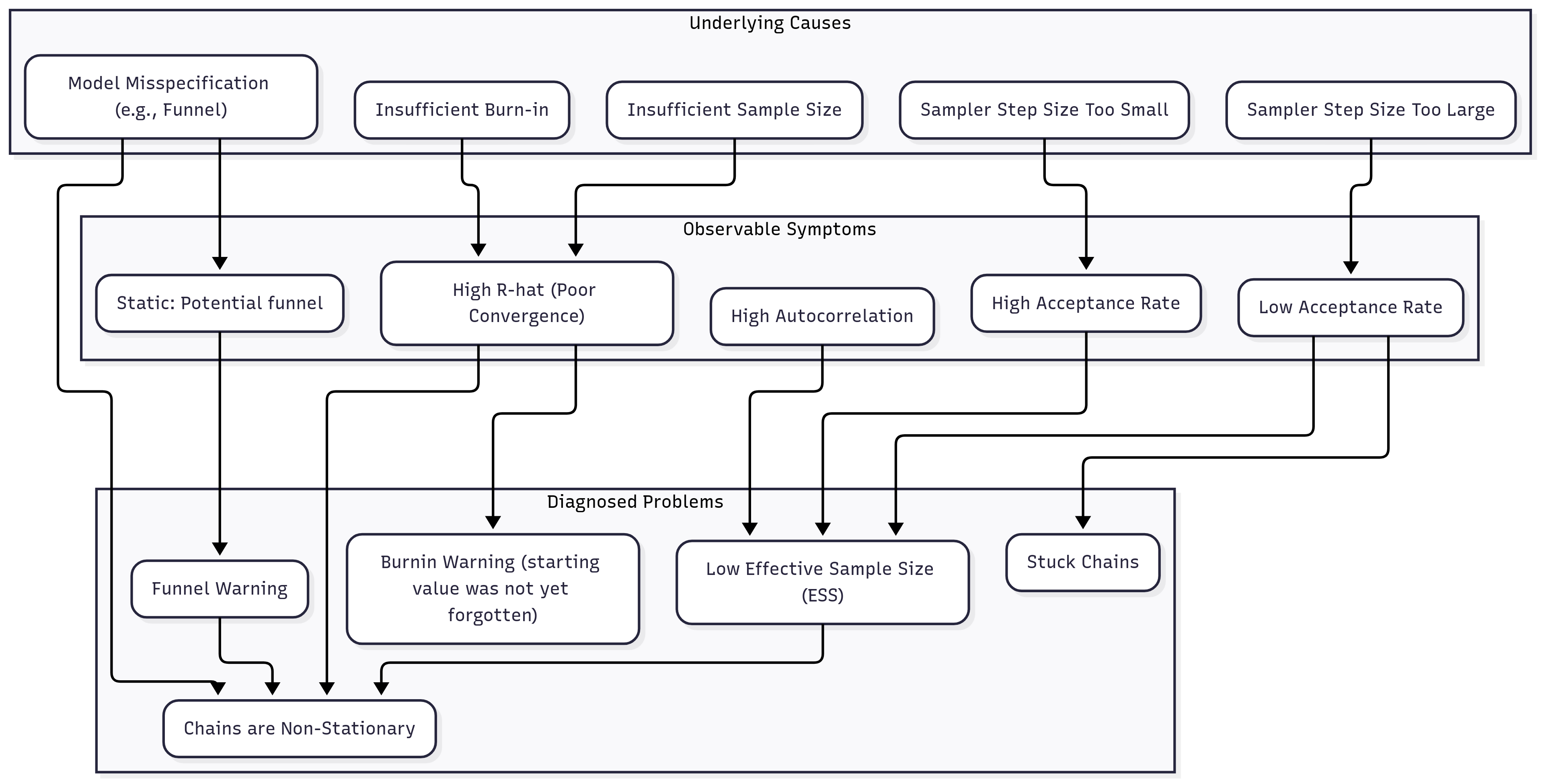}
\caption{Warnings Relation. \textmd{Showing the causes, observable values and triggered warnings.}}
\label{fig:warnings_graph}

\end{figure}

While interactivity already leads to a less cumbersome analysis experience, it still does not reduce the amount of steps taken to detect an inference issue.
As stated in \textit{DI3}, the number of configurations for a single analysis step (selection of variables, chain, and diagnostic tool) is very large even for simple models.
Therefore, it is critical for a debugger to implement heuristics and automated checks that guide the user to subsets of this debugging space which are most likely to reveal issues in the inference result.

\begin{figure}[h!]
\centering
\includegraphics[width=0.4\textwidth,trim={0 0 0 1cm}]{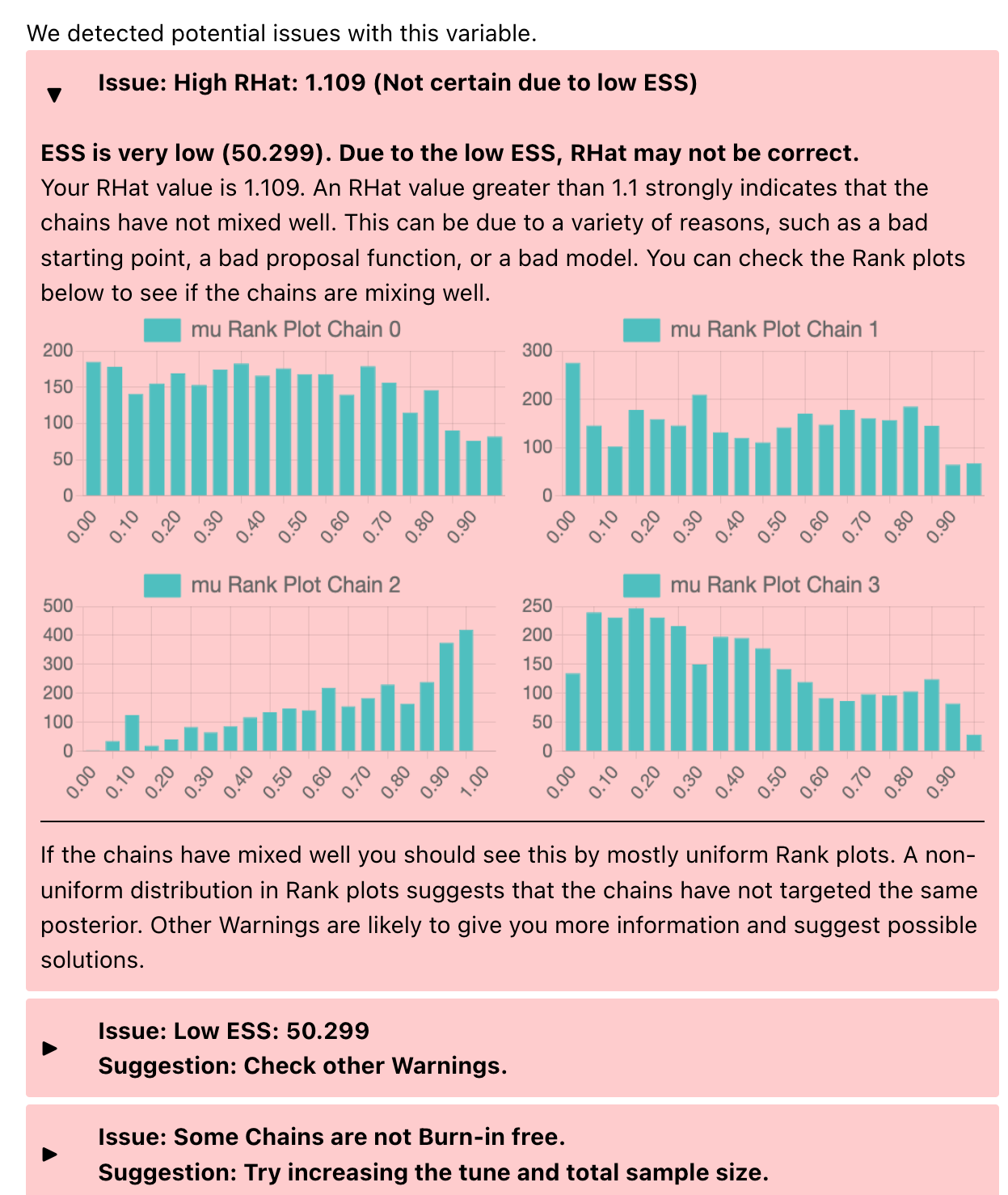}
\caption{Example of warnings raised by the debugger. \textmd{This example shows one of the debuggers warnings in detail and 2 others collapsed. The warnings try to highlight probable issues and aim to inform the user to a point were they can make a decision if an action is needed or not. Whenever possible warnings show a suggestion that hopefully fixes the underlying issue.}}
\label{fig:warnings_example}
\end{figure}

\FloatBarrier

\begin{table}[h!]
\begin{tabular}{p{1.2cm}|p{1.8cm}|p{2.8cm}|p{1.8cm}}
\hline
\textbf{Warning} & \textbf{Key Symptoms} & \textbf{Likely Cause(s)} & \textbf{Suggested Solution (from code)} \\ \hline
High R\textasciicircum{} & $\hat r$ \textgreater{} 1.01 & Chains have not converged to the same distribution / Multiple & "See other warnings." Check rank plots. \\ \hline
Burn-in Warning & High $\hat r$ in the first part of a chain. & Initial samples are not from the posterior distribution. & "Increase the burn-in period." \\ \hline
Funnel Detected \& Low/High Acceptance & A parameter's variance is controlled by another parameter. & Model parameterization causes posterior that is difficult to sample from. & "Reparameterize the model." (Provides code) \\ \hline
Low ESS \& High Acceptance & ESS is low AND acceptanceRate is high. & Sampler step size is too small, leading to high autocorrelation. & "Increase the proposer's step size." \\ \hline
Low ESS \& Low Acceptance & ESS is low AND acceptanceRate is low. & Sampler step size is too large, leading to high rejection rates. & "Lower the proposer's step size." \\ \hline
Stuck Chain & A chain has not accepted a new sample in many iterations. & Extremely low acceptance rate. & "Check your proposal functions and step size." \\ \hline
Low ESS (Isolated) & ESS is low. & High sample autocorrelation, reason unclear. & "Check other warnings, they might be indicative." \\ \hline
Low/High Acceptance (Isolated) & acceptanceRate is outside the optimal range. & Step size is likely miscalibrated, but effect on efficiency is not yet clear. & "Maybe change the step size." \\ \hline
\end{tabular}   
\caption{Warnings and their triggers. \textmd{Shows all the possible warnings raised by \toolname{} together with a brief overview of triggers, likely causes and suggestions.}}
\label{table:warnings}
\end{table}

We realized this requirement in \toolname{} in the \emph{Warnings View}.
In the background, \toolname{} monitors a set of metrics relevant to MCMC inference including acceptance rate, effective sample size, and the rank-normalized $\hat r$.
Then, a number of analyses are run on these metrics and the \emph{Warnings View} is populated with messages that warn the user about instances indicative of faulty inference along with suggestions on how to fix them. These checks include static analyses that run on the source code of the program, which we have implemented with the LASAPP framework~\cite{boeck2024lasapp}.
For a complete overview of the implemented warnings see \Cref{table:warnings}. \newText{Fig~\ref{fig:warnings_graph} further shows the relationship between symptoms, analysis and warnings.}

\begin{figure*}[h!]
\includegraphics[width=0.9\textwidth]{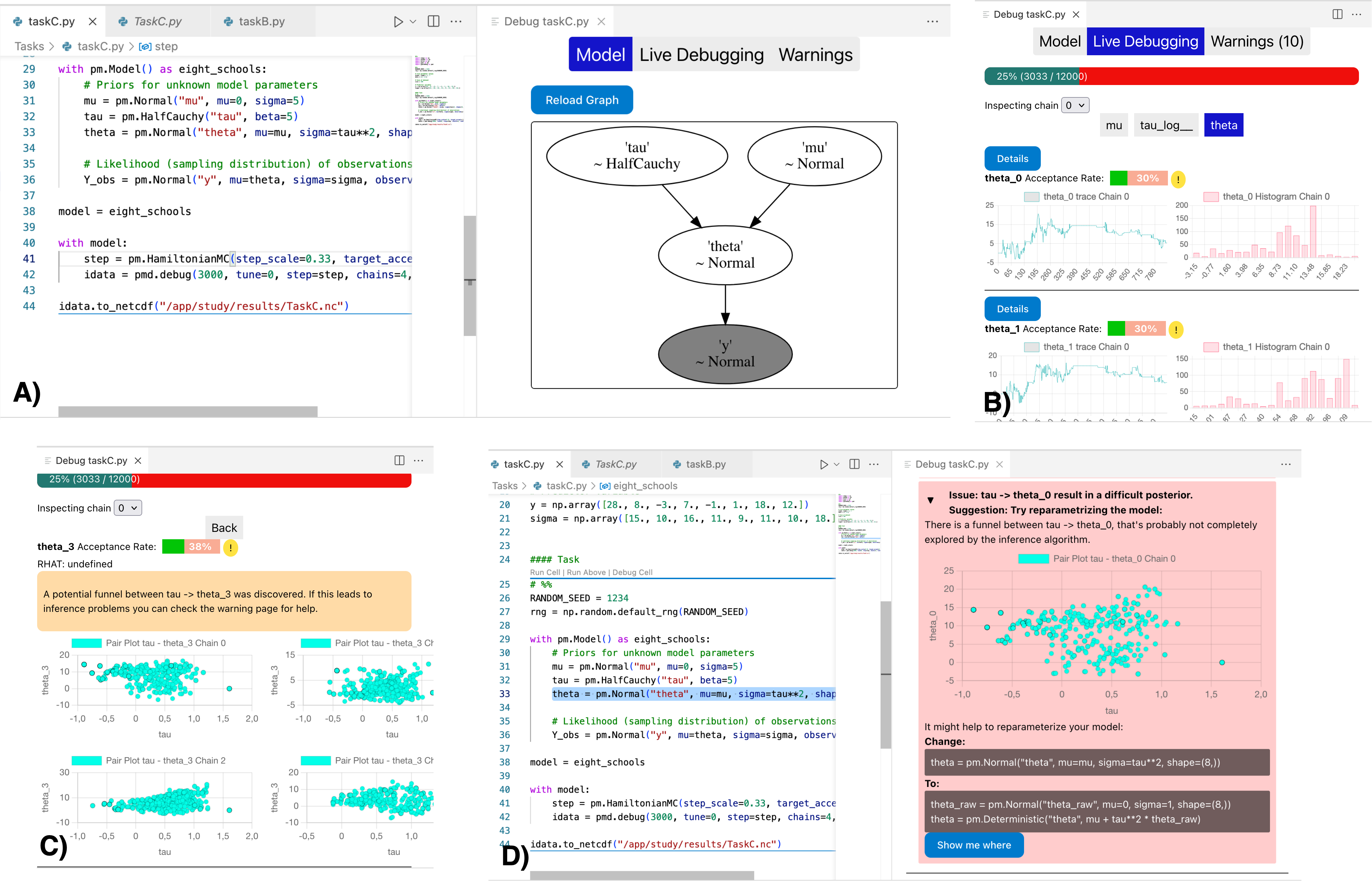}
\caption{Exemplified workflows. \textmd{\textbf{A)} shows the model graph view of the tool right next to the source code. \textbf{B)} is the \textit{Live Debugging View}, which gives insights into the developing posteriors. \textbf{C)} shows the \textit{Details View} for variable \textit{theta\_3} with included pair plots due to the statically identified variance dependence between \textit{tau} and \textit{theta}. In \textbf{D)} we see a warning with explanation, graph and the identified line in the source code that causes the issue, together with a suggestion for a solution.}}
\label{fig:example_workflow}
\end{figure*}

\subsection{REQ6: Easy Integration with Existing PPLs}
Lastly, it is important that our debugger can easily be applied to existing PPLs.
This requires a light-weight API to send data during inference to the debugging and analysis backend.

After sending initial information, \toolname{} only requires the PPL backend to make \newText{regular HTTP requests, either batched or after each MCMC iteration, sending the current inference state as JSON.}\delete{one HTTP request after each MCMC iteration sending the current inference state as JSON.}
Since PyMC provides an API for making such callbacks, the integration was particularly easy. \newText{Some other PPLs offer similar APIs, but even if a PPL does not, the HTTP calls could be directly implemented in the language without significant challenges. By implementing the HTTP interface within a language, most features of \toolname{} are immediately available. To get full support for contextualized workflows, the \textit{ConcretePPL} interface inside \toolname{} must be implemented for a new language. This allows \toolname{} to add additional information to the warnings that are specific to the given PPL.}

\section{Workflows in \toolname{} Exemplified}

To illustrate the accelerated workflows enabled by \toolname{}, let us revisit the SAT coaching example from \Cref{sec:debugging-insights} from the users perspective.

\paragraph{Modeling} Suppose a user wants to investigate whether short-term coaching improves SAT scores, using a Bayesian model. They begin implementing a PyMC model, informed by their prior beliefs. At some point, they want to see how their model is structured. This is when they open \toolname{}, which shows the model code side-by-side with a visual representation of the model graph (Fig.~\ref{fig:example_workflow} Step~\textbf{A}).

\paragraph{Live Feedback During Inference} When ready, they launch inference. Immediately, live diagnostics appear (Fig.~\ref{fig:example_workflow} Step~\textbf{B}). As posterior samples stream in, the user can inspect evolving distributions, trace plots, and summary statistics in real-time. The user starts inspecting the progression of the different variables. When they look at an overview of the $\theta$ variables the user notices a warning icon next to the acceptance rate, which is at only 30\%. 

\paragraph{Focused Inspection} Since the acceptance rate is very low for \textit{HMC} based algorithms and the traceplots are indicative of suboptimal convergence, the user opens the details view for one of the $\theta$ variables (Fig.~\ref{fig:example_workflow} Step~\textbf{C}). The user immediately notices the orange text box that tells them that there is a potential funnel between $\tau$ and $\theta$\newText{, a shape that is often difficult to explore for MCMC based algorithms}. This leads them to inspecting the shapes in the conditionally displayed pairplots between these two variables, which indeed look funnel-shaped.

\paragraph{Warnings and Potential Solutions} While the user is inspecting the details page, they notice that there are 10 warnings for potential problems discovered by \toolname{}. So the user heads over to the warnings view and opens the topmost warning (Fig.~\ref{fig:example_workflow} Step~\textbf{D}) for the $\theta_0$ variable. This warning, again, raises the issue of the funnel between $\tau$ and $\theta$ and shows that the resulting posterior is difficult for \textit{HMC} algorithms to explore. It further includes the problematic line of code together with a suggested change. The user accepts this suggestion and clicks on the button "Show~me~where" to jump to the exact line of code causing the issue and replaces it.

\paragraph{Iterations} Since continuing inference after identifying a fault wastes time, the user stops inference early and restarts it with the applied changes in place. From there the user keeps iterating together with \toolname{} until all warnings are resolved and the inference results show no indication of non-convergence.

\section{Study Design}

To evaluate our framework we conducted a study with 18 participants.
We formulate following research questions:
\begin{enumerate}
    \item [(RQ1)] Does our online and interactive framework effectively decrease time spent debugging inference?
    \item [(RQ2)] Are online visualizations of MCMC posteriors considered useful by PPL users?
    \item [(RQ3)] Are the warnings raised at inference time considered useful by PPL users?
\end{enumerate}

To answer these questions each participant was asked to solve three inference debugging tasks in PyMC and to fill out two questionnaires: one pre-study questionnaire and one post-study questionnaire. 


RQ1 is answered quantitatively by evaluating the hypotheses stated in \Cref{tab:hypothesis} with statistical tests.

\begingroup

\setlength{\tabcolsep}{4pt} 
\renewcommand{\arraystretch}{1.4} 

\begin{table}[h!]
    \centering
    \resizebox{0.8\columnwidth}{!}{
    \begin{tabular}{p{8mm}p{3cm}|p{6mm}p{3cm}}
        \hline
        & Null hypothesis & & Alternative hypothesis \\
         \hline
        $H1_0$ : & The tool does not impact how many issues PPL user can solve & $H1$ : & PPL users are able to resolve more issues with the tool \\

        $H2_0$ : & The tool does not impact how fast PPL user can identify issues & $H2$ : & PPL users are able to identify issues faster with the tool \\

        $H3_0$ : & The tool does not impact how fast PPL users iterate over the problem & $H3$ : & PPL users have faster iteration cycles with the tool \\

        $H4_0$ : & The tool does not impact how effective PPL users spend their time & $H4$ : & PPL users spend their time more effectively with the tool \\
        \hline
    \end{tabular}
    }
    \caption{Null and alternative hypothesis}
    \label{tab:hypothesis}
\end{table}

\endgroup

For \textit{RQ2} and \textit{RQ3} we adopt a qualitative methodology, basing our analysis on participants responses to the questionnaires and verbal statements they made while performing the tasks.

\subsection{Study Setup}
The study was conducted as a controlled experiment with a duration of 1h:15m to 2 hours for each participant.

\paragraph{Environment:} The study was conducted fully online via a video conferencing tool.
To avoid differences in inference runtime depending on hardware differences, we provided an online version of VSCode.
The participants accessed the online IDE via \url{https://vscode.dev} and all program code was executed on the same lab-server for every participant.

\newText{
\paragraph{Design \& Baseline Condition:} The study follows a within-subject design, where the task order was randomized and the first task serves as the \emph{baseline} condition for each participant.
For this baseline, participants were allowed to use any existing tool or resource of their liking, \emph{except} \toolname{}, to closely reflect real-world workflows.
With Task 2 and 3 we evaluated the \emph{tool} condition, where the participants were allowed to use \toolname{}
\emph{in addition} to existing tools or resources.

For the first task, we observed participants opting to use the following existing tools and resources during: prior and posterior predictive checks, ArviZ (plots + summary diagnostics), search engines, PyMC + ArviZ documentation and tutorials, ChatGPT. In general, we observed that participants had different workflows but often relied heavily on ArviZ and search engines. Interestingly, only one participant asked a large language model for help. 
Across all participants, \toolname{} became the primary tool during Tasks 2 and 3, with many relying on it so extensively that they stopped using other tools and resources.
}

\paragraph{Protocol:}
Before the experiment, participants were asked to fill-out a pre-study questionnaire to assess their background and familiarity with probabilistic programming.
Afterwards the participants connected to the online environment and were given an introduction notebook. 
Its main purpose was introducing participants to PyMC and relevant tooling around PyMC. Participants were told to take as much time as they needed and that they could always refer back to the notebook or use any other resource they wish. 

Once a participant gave the signal that they were ready, they could proceed to the first task. The participants were informed that they have to solve the tasks in order and cannot go back and forth between them. The task order was randomly assigned to each participant, and every participant had to solve the first task without \toolname{}.
However, they could use any tool or resource they had access to for all tasks.
The time limit for each task was set to 30 minutes (subtracting setup issues), and participants had to decide for themselves when a task is finished. They did not receive any feedback on their progress during the study. After participants deemed the first task done (or reached the time limit) they received a $\sim$5 minutes introduction to \toolname{}. Participants were now free to use \toolname{}, if they wanted, for the following two tasks. Aside from the new tool, everything else stayed the same. Once the participant finished all three tasks, they were asked to fill-out a post-study questionnaire. Afterwards, the study concluded and participants received feedback on their task performance if they wished.

\subsection{Tasks}

A task in our study is defined by a probabilistic program given as a PyMC source file with pre-implemented model and inference routine.
In these programs we have introduced sub-optimal or faulty implementation decisions.
The objective for the participants is to determine whether there is an issue with the model and/or inference implementation or whether there is no issues at all.

We have prepared three tasks in total:
\begin{enumerate}[label=(\Alph*)]
    \item Linear regression model with HMC inference routine. \textit{Fault:} bad choice of step-size hyper-parameter of HMC.
    \item Signal detection theory model \cite{lee2014bayesiancognitivemodelling} with HMC inference routine. \textit{Fault:} Unused random variables, sub-optimal hyper-parameter choice for step-size and too short burn-in period and too small sample size.
    \item Eight schools model \cite{Rubin1981EstimationExperiments} with HMC inference routine. \textit{Fault:} Sub-optimal model parameterization, too short burn-in period and depending on burn-in period setting potentially auto-tune needs to be disabled.
\end{enumerate}
The tasks vary in difficulty: the simple model and poor inference result of task A makes it easier to identify and solve in comparison to the other more challenging tasks B and C.

\subsection{Participants}
Probabilistic programming is a relatively small community and since our study targets \textit{PPL} users and requires prior knowledge of Bayesian inference preferably in the context of probabilistic programming, recruiting participants became a challenging task.


As our efforts for recruiting study participants in the forums of the three popular PPLs PyMC, Pyro, and Stan was unfortunately unsuccessful, we opted for a \textit{snowball sampling} \cite{Atkinson2001AccessingStrategies} strategy instead.
Here, we considered three major pools:
\begin{enumerate}
    \item Students who took a university course on Probabilistic Programming \emph{(8 participants recruited)}
    \item Probabilistic Programming and Bayesian modeling researchers \emph{(7 participants recruited)}
    \item Practitioners, e.g., other scientists, data analysts, etc. \emph{(3 participants recruited)}
\end{enumerate}


\subsection{Result Validation and Statistical Tests}
To make our results verifiable, we conducted a series of  statistical tests, where we set the significance level at $5\%$.

\subsubsection{Ordinal Measurements of Uncensored Data}
Due to the non-normality of our ordinal variables (confirmed with the Shapiro-Wilk test), we performed the Mann-Whitney U Test on these measurements.
The Mann-Whitney U Test is non-parametric and thus makes no assumption on the distribution of our data.
Following common practice, we further used Cliff's delta for estimating the effect size~\cite{CitoInteractiveIDE, SalvaneschiIEEEStudy}.

\subsubsection{Censored Data}

Some of our data variables are \emph{right censored}, which means that there may be events that did not occur within the study period.
For instance, the \textit{Time to first issue detected} is right-censored, because some participants were unable to detect the first issue within the given time limit.
For these kinds of variables, statistical survival analysis can be adopted~\cite{Bewick2004StatisticsAnalysis,lee2003statisticalSurvival}.
We performed \textit{Cox Proportional Hazards Regression}~\cite{CoxPropHazard} and a log-rank test~\cite{mantel1966evaluation} to estimate the effect of our framework on the "survival rate".
In this context, survival refers to the event that a bug "survives" the detection efforts of the participant.

Since our study measures a positive outcome (e.g. task solved) in contrast to many other studies using these methods, we will focus our attention on the cumulative densities $cd(T)$, instead of the survival function $sf(T)$, which is defined as $cd(T) = 1 - sf(T)$.
So instead of investigating the probability of a task remaining incomplete beyond time $T$ (e.g., survival), we focus on the probability of the task being completed (commonly interpreted as death in survival analysis) prior to time $T$.

\subsection{Threats to Validity}

\subsubsection{External Validity}
Given that participation in a study is entirely voluntary and the participants received no material benefit from participation, it is possible that our sample is not representative of the general population of \textit{PPL} users. Furthermore, the use of snowball sampling may lead to a more homogeneous group of participants \cite{CitoInteractiveIDE}. We mitigated these issues by reaching out to possible participants broadly and ensuring that our participants come from different backgrounds. We were especially careful, to not only recruit students but also researchers and professionals to have participants from the full spectrum of \textit{PPL} users. \newText{Furthermore, we ensured that the recruited participants had no dependency on the researchers (e.g. all students have been graded before participating).}

\subsubsection{Internal Validity}
We identified two main threats to internal validity for this study. First, due to the necessarily artificial nature of the tasks and setup, it is possible that the tasks do not represent real world issues well. Second, the tasks may overly favor the usage of \toolname{}, through aligning with its feature set.
We mitigated both these threats through basing our examples on existing literature and known examples. Task~C, for example, is a well known problem in probabilistic programming research, Task~B is based on a tutorial task for the \textit{Pyro PPL}~\cite{bingham2019pyro} with the main issue being a copy\&paste / typo, which was an actual mistake made during task creation and evaluation and Task~A is the prototypical linear regression example with a common issue of bad hyper-parameters. \newText{To further mitigate the effect of task alignment, we introduced multiple issues into Task~B and Task~C.
For instance, while the funnel warning in \toolname{} was designed to detect funnel relationships as constructed by the centered eight school model in Task~C, we introduced two additional issues in the program to reduce alignment.
Nevertheless, it is possible that the effects measured in this study, are somewhat larger than they would be in real world scenarios with more subtle bugs.}

\section{Quantitative Results}

In this section, we present our quantitative analysis to answer the question \textit{RQ1: Does our online and interactive framework effectively decrease time spent debugging inference?}
In the following subsections, we show that we can reject each of the four null hypotheses of \Cref{tab:hypothesis}.

\subsection{H1: Users are able to resolve more issues}

Table \ref{table:avg_percent_solved} shows the mean and standard deviation of the percentage of resolved issues across all tasks, together with p-values from Mann-Whitney U test and Cliff's delta.

\begin{table}[h!]
\begin{tabular}{lllrr}
\toprule
Task & Tool & Baseline & Cliff's $\delta$ & p-value \\
\midrule
A & 91.67 ± 28.87 & 66.67 ± 51.64 & 0.250 & 0.218 \\
B & 69.45 ± 26.43 & 5.55 ± 13.61 & 0.958 & \textbf{0.001} \\
C & 75.00 ± 32.18 & 5.55 ± 13.61 & 0.889 & \textbf{0.002} \\
\bottomrule
\end{tabular}
\caption{Average Percent of solved issues (Mann-Whitney U).}
\label{table:avg_percent_solved}
\end{table}
The results show a significant difference between the groups for the tasks \textit{B} and \textit{C}.
For the simplest task \textit{A}, we could not find a significant difference between the groups.
This suggests that our framework provides greater help as tasks grow in complexity and potentially involve multiple issues. Considering that task \textit{A} only involves 3 latent variables whereas task \textit{C} involves 10 and task \textit{B} 20, this shows that our tool is likely to reduce the exploration space for larger problems. This further suggests that our tool successfully tackles the previously examined debugging insight \textit{DI3}, where we discussed the exploration space quickly growing as the number of variables to consider increases.


For hypothesis H1, we further conducted a Bayesian analysis of the results, which allows us to quantify effects in relation to each other, in contrast to the standard statistical tests.
We implemented a Bayesian logistic regression model, which represents the number of issues identified as a Bernoulli distribution.
The effects are modeled as latent random variables and are summed up and passed through a sigmoid function, which outputs the probability $p$ of the Bernoulli distribution.

\begin{figure}[h!]
\centering
\includegraphics[width=0.8\columnwidth]{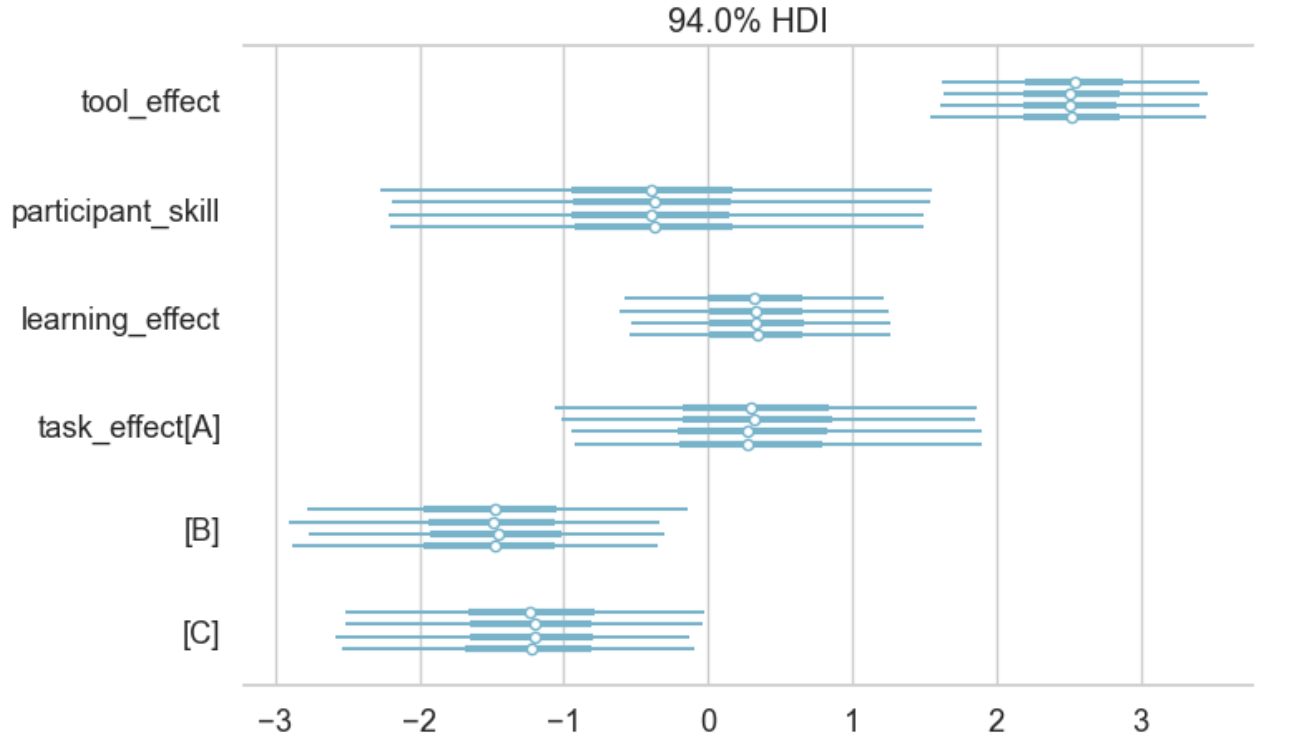}
\caption{Forest plot of effect posteriors from Bayesian analysis.}
\label{fig:bayesian_effect}
\end{figure}

Fig~\ref{fig:bayesian_effect} shows a forest plot of the posteriors for \textit{tool effect}, \textit{participant skill}, \textit{learning effect}, and \textit{task effects}. From the plot we see a clear indication of \toolname{} having a large positive effect on the probability of solving an issue. 

\begin{figure}[h!]
\centering
\includegraphics[width=0.5\textwidth]{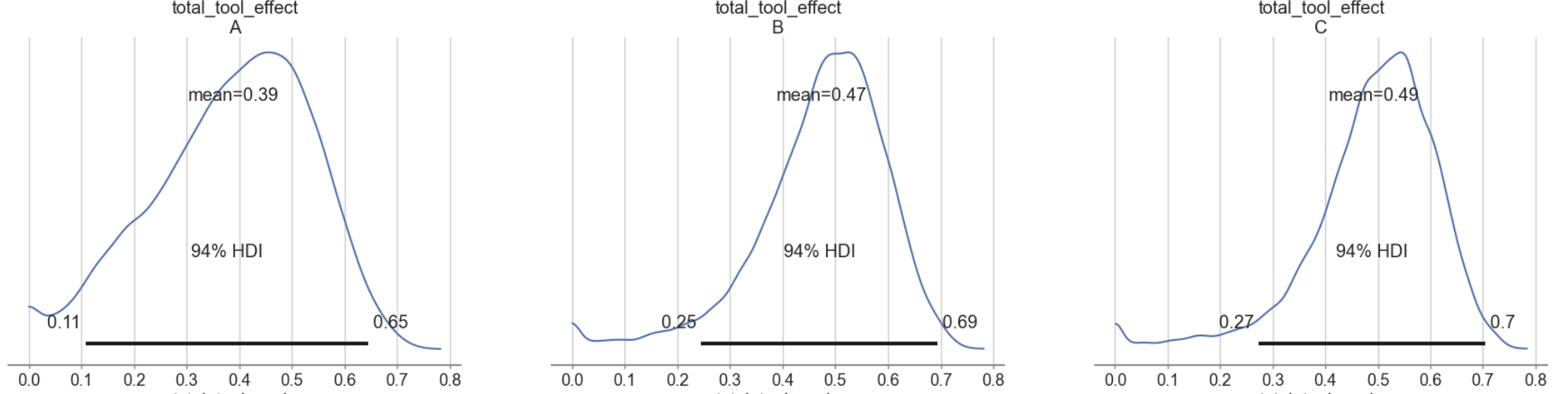}
\caption{Distribution over effective increase in percentage points for solving any issue if the tool is used, compared to the tool not used.}
\label{fig:bayesian_effect_increase}
\end{figure}

We see that on average, the usage of the tool increases the probability of solving an issue by 39\% points, 47\% points, and 49\% points for tasks A, B, and C, respectively. This means the average user has a 20\% chance of solving any given issue in Task C without the tool and a 69\% chance with \toolname{}. The distributions for these increased probabilities can be seen in Fig.~\ref{fig:bayesian_effect_increase}.

More detailed results and descriptions of the implemented model can be found in the online appendix to this paper (\url{https://github.com/ipa-lab/InferlogHolmes-Appendix}).

\begin{table*}[h!]
    \centering
    \resizebox{\textwidth}{!}{%
    \begin{tabular}{llrrrrrrrr|r}
        \toprule
         & Task & coef & exp(coef) & coef lower 95\% & coef upper 95\% & Concordance & Log-Likelihood ratio test & z & CoxPH p & Log-Rank p \\
        covariate &  & &  &  &  &  &  &  & & \\
        \midrule
        tool & A & 1.9640 & 7.1278 & 0.5353 & 3.3927 & 0.70 & 9.65 on 1 df & 2.6944 & \textbf{0.0071} & \textbf{0.0016} \\
        tool & B & 2.2097 & 9.1134 & 0.7922 & 3.6273 & 0.74 & 12.31 on 1 df & 3.0553 & \textbf{0.0022} & \textbf{0.0003} \\
        tool & C & 2.3902 & 10.9160 & 0.6806 & 4.0998 & 0.74 & 11.19 on 1 df & 2.7403 & \textbf{0.0061} & \textbf{0.0007} \\
        tool & All Tasks pooled & 3.4129 & 30.3523 & 1.9264 & 4.8993 & 0.72 & 45.93 on 1 df & 4.5000 & \textbf{<0.0000} & \textbf{<0.0000} \\
    \bottomrule
    \end{tabular}
    }
    \caption{Time to First Issue Detected depending on tool used Cox-Proportional-Hazard and Log-Rank test results. \textmd{Results from the CoxPH test. \textit{exp(coef)} shows how much the tool influences the hazard rate.}}
    \label{table:first_issue_log_rank}
\end{table*}

\subsection{H2: Users are able to identify issues faster}

Table \ref{table:first_issue_log_rank} shows the results from Cox-Proportional-Hazard regression analysis depending on tool use and for comparison p-values from a Log-Rank test for each task individually and for all tasks pooled. \newText{To interpret the results of the Cox Proportional-Hazards regression analysis, we examine the \textit{exp(coef)}, which represents the estimated hazard ratio produced by the model. For a more comprehensive discussion of the interpretation of survival analysis methods, we refer the interested reader to \cite{gareth2023introduction}.} Due to the small number of occurred events during the task time, in some groups the results should be taken with care. Nevertheless, the large \newText{\textit{exp(coef)}} coefficient together with small p-values suggests that these differences would be highly unlikely under the null-hypothesis of equality between the groups. Furthermore, when we pool all tasks together, we see an even stronger effect, and the data suggests that a user has a \textbf{30x} higher chance of discovering an issue at any given point in time with our tool than without.

\begin{figure}[h!]
\includegraphics[width=0.8\columnwidth]{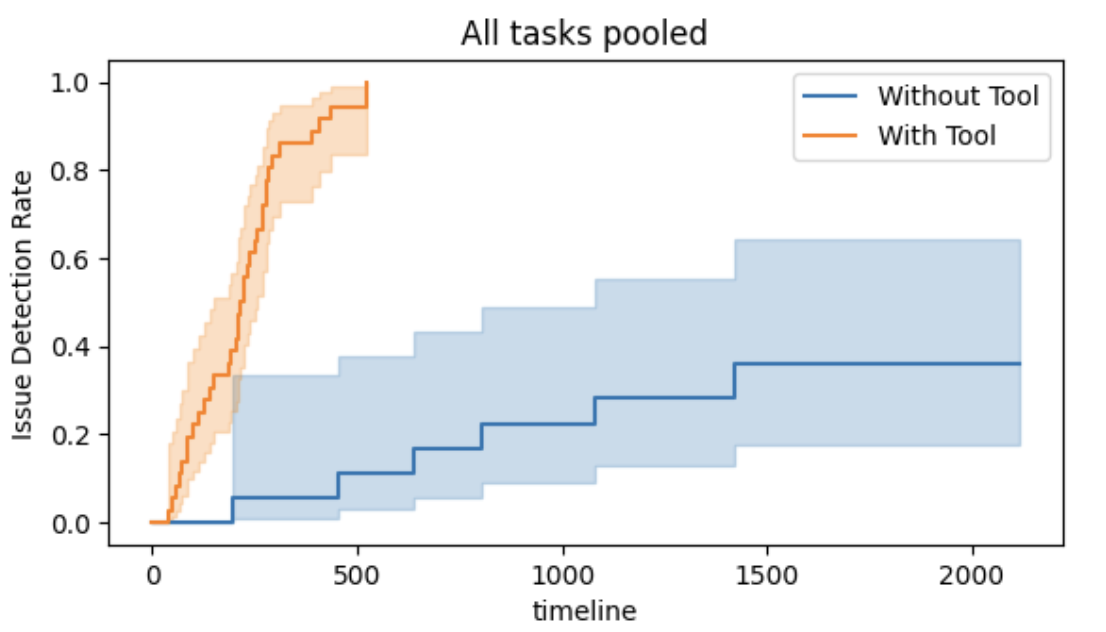}
\caption{Time to First Issue Detected (seconds) Kaplan-Meier-Estimator Cumulative Densities Curves for all tasks pooled.}
\label{fig:first_issue_detected_KM}
\end{figure}

These results are supported by visual inspection of the Kaplan-Meier-Estimator's Cumulative Density Curve~\cite{Kaplan01061958} in Fig.~\ref{fig:first_issue_detected_KM}. With our developed framework, there is an \textbf{83.3\%} chance for a user having detected an issue within the first 5 minutes. Without our tool the same chance is only \textbf{5.6}\%.

\subsection{H3: Users have faster iteration cycles} Table \ref{table:iterations_per_task} shows the mean and standard deviation of three measurements together with p-values from Mann-Whitney U Test and Cliff's delta. \textbf{Inference Canceled} describes the number of times a participant ran inference but actively canceled it before it finished. \textbf{Inference Started} is the number of times a participant ran inference on the task, and \textbf{Changes Made} represents the number of times a participant changed either parts of the model or of the inference algorithm.

\begin{table}[h!]
    \resizebox{0.45\textwidth}{!}{%
    \begin{tabular}{llllrr}
        \toprule
        Task & Measurement & Tool & Baseline & Cliff's $\delta$ & p-value \\
        \midrule
        \multirow{3}{*}{A} & \# Inference Canceled & 1.33 ± 1.44 & 0.17 ± 0.41 & 0.625 & \textbf{0.026} \\
         & \# Inference Started & 4.33 ± 1.72 & 2.50 ± 0.84 & 0.667 & \textbf{0.024} \\
         & \# Changes Made & 3.08 ± 1.88 & 2.17 ± 1.94 & 0.292 & 0.341 \\
        \hline
        \multirow{3}{*}{B} & \# Inference Canceled & 1.08 ± 1.00 & 0.17 ± 0.41 & 0.556 & \textbf{0.048} \\
         & \# Inference Started & 4.25 ± 1.06 & 2.33 ± 1.03 & 0.792 & \textbf{0.007} \\
         & \# Changes Made & 3.92 ± 1.31 & 1.50 ± 1.38 & 0.778 & \textbf{0.009} \\
         \hline
        \multirow{3}{*}{C} & \# Inference Canceled & 2.42 ± 2.39 & 0.17 ± 0.41 & 0.736 & \textbf{0.010} \\
         & \# Inference Started & 5.75 ± 2.42 & 2.33 ± 1.37 & 0.875 & \textbf{0.003} \\
         & \# Changes Made & 4.75 ± 2.09 & 1.67 ± 1.21 & 0.847 & \textbf{0.004} \\
        \bottomrule
    \end{tabular}
    }

    \caption{Comparing iterations with and without tool (Mann-Whitney U). \textmd{The table compares results from the Mann-Whitney U test for all 3 tasks on 3 different indicators for iteration speed. \textit{Inference Canceled} refers to how often participants stopped inference early. \textit{Inference Started} refers to how often participants started the inference routine, and \textit{Changes Made} shows the number of changes made to the model or the inference algorithm.}}
    \label{table:iterations_per_task}
\end{table}

With the exception of \textit{number of changes} performed at task \textit{A}, we see significant differences across all measurements. Especially the number of times participants ran inference and the number of times they canceled inference early suggest that our framework accelerates the probabilistic programming workflow by facilitating more and faster iterations. For \textit{B} and \textit{C}, we also see a significant increase in changes made by participants. One possible explanation why we do not see this increase for task \textit{A} is that task \textit{A} is solvable in at least three different ways with only a single change. Task \textit{B} and \textit{C} in contrast have no single change solution (to our knowledge) and both require multiple changes to be fully solved.

\subsection{H4: Users spend their time more effectively}

Table \ref{table:percent_of_time_inference} shows the mean and standard deviation of the percentage of time spent in inference relative to the total time spent across all tasks, together with p-values from Mann-Whitney U Test and Cliff's delta.
A high percentage is indicative of a participant being able to debug issues while inference was running instead of having to wait for inference to complete and debug afterwards -- the typical debugging experience in probabilistic programming before \toolname{}.

\begin{table}[h!]
\begin{tabular}{lllrr}
\toprule
Task & Tool & Baseline & Cliff's $\delta$ & p-value \\
\midrule
A & 0.54 ± 0.11 & 0.23 ± 0.10 & 0.944 & \textbf{<0.001} \\
B & 0.56 ± 0.11 & 0.23 ± 0.09 & 1.000 & \textbf{<0.001} \\
C & 0.61 ± 0.13 & 0.27 ± 0.15 & 0.917 & \textbf{0.001} \\
\bottomrule
\end{tabular}
\caption{Average percent of time spent in inference relative to total time spent on task(Mann-Whitney U).}
\label{table:percent_of_time_inference}
\end{table}

The results suggest that, indeed, participants have reduced waiting times when using \toolname{} and that they can use the time spent in inference effectively for analyzing and debugging problems.
Tasks they would otherwise do after inference completed, can now be done at inference time, which suggests better time utilization.


\section{Qualitative Analysis and Participants' Perception}

In this section, we present a qualitative analysis of participants' perceptions to answer \textit{RQ2} and \textit{RQ3}. We base our analysis on responses to the post-study survey and quotes from participants recorded during the study.

\subsection{RQ2: Are online visualizations of MCMC posteriors considered useful by PPL users?}

To answer RQ2, we tried to separate the effect of online visualizations and visualizations in general. We designed our questionnaire to differentiate between the selection of visualizations in the tool and the general idea of online visualizations during inference.

\subsubsection{The tool provides a good selection of visualizations}
\begin{figure}[h!]
\centering
\includegraphics[width=0.5\textwidth]{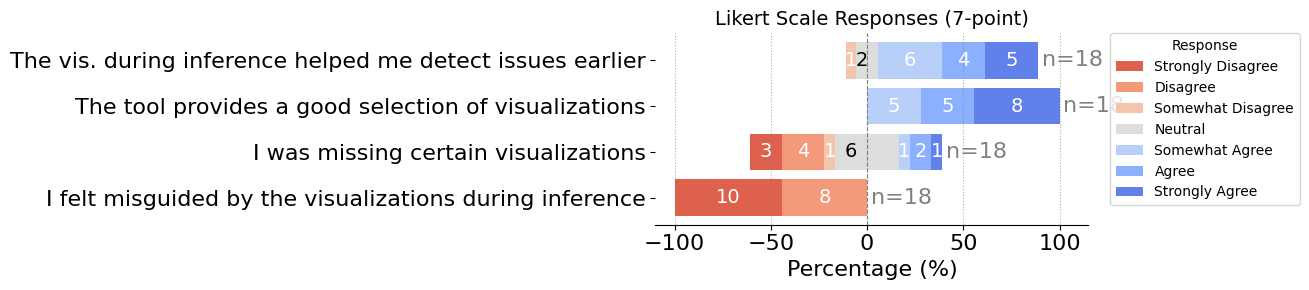}
\caption{Visualization responses}
\label{fig:viz}
\end{figure}

As can be seen in Fig.~\ref{fig:viz}, participants considered the selection of visualizations overwhelmingly positive and only a few participants were missing certain visualizations.

One participant mentioned that they missed a clean overview of posterior statistics like Arviz \cite{Kumar2019ArviZPython} provides through the \textit{summary} function. Another participant related their statement about missing visualizations towards the static model graph. They mentioned that they are "[...] always missing the formula how parameters are added or mutiplied in the graph view as well in the tools case also shape information". 

In another section of the questionnaire one participant made the remark that they "[...] would like for each visualization to have an explanation of on how to read it in detail, maybe via a link to another page". 

We consider these points valid feedback and incorporating them into a future version of the tool should be straight forward without negatively impacting the tools usefulness.

\subsubsection{Online visualizations help detect issues faster}

We further see in Fig~\ref{fig:viz} that almost all participants found the visualizations during inference helpful for issue detection. With only 1 participant somewhat disagreeing and 2 answering neutral. One participant said during their third task: "It is nice to see how the chain develops" and another exclaimed at their second task: "It's actually really nice to see it happening". A third participant compared the tool to tensorboard \cite{MartinAbadi2015Systems} and called it "very comfortable". One specifically highlighted at the post-study survey, that they "[...] really like not having to wait for the model to be finished especially as pymc model[s] have long runtimes". The same participant also pointed out that the tool helped them "[...] learn new debugging visualizations [they] wouldnt use otherwise as [they] also have to create them which takes time".



\subsection{RQ3: Are the warnings raised at inference time considered useful by PPL users?}

As before, we tried to separate the perception of the warnings in general and the perception of the warning during inference.  

\subsubsection{In general, warnings are considered helpful}
\begin{figure}[h!]
\centering
\includegraphics[width=0.5\textwidth]{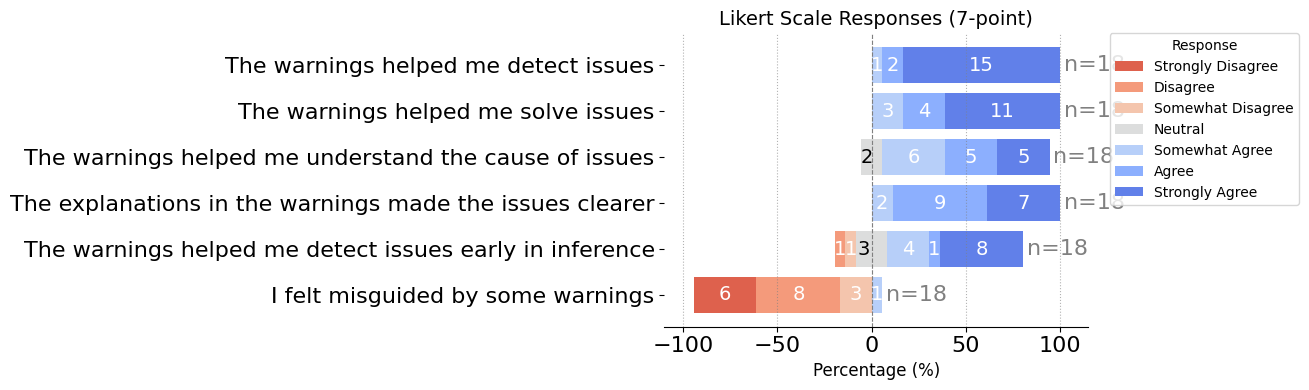}
\caption{Warning responses}
\label{fig:warnings}
\end{figure}

From the responses presented in Fig.~\ref{fig:warnings}, we see that a clear majority of participants strongly agreed that warnings helped them detect issues and only 2 participants "just" agreed and 1 participant somewhat agreed with the statement. While the statement "The warnings helped me solve issues" drew less strong agrees to it, the overall tendency is that the warnings are helpful in that regard as well.

This sentiment is further supported when participants were asked, if the warnings helped them understand issues and if the warnings made issues clearer. In both cases the majority of participants either agreed or strongly agreed with the statement. One participant highlighted in the optional free form input field of the post study questionnaire that "they[the warnings] were quite informative [...]". Another mentioned that the warnings are "very helpful with nice explanations!" and one participant pointed out that "[...] they were very helpfull[sic!], as it is hard to find from the docs, which parameters might influence the behaviour in a certain way". During their third task (task \textit{C}) one participant exclaimed: "oh that's cool. Well let's try it out, when the tool says that". This was subjected at the funnel/reparametrization warning described in Table~\ref{table:warnings}.

\subsubsection{Warnings are considered mostly helpful during inference}
While the helpfulness of warnings during inference is still perceived positive overall, participants agreed less strongly to it than to the general helpfulness of the warnings. One participant even disagreed, and another one somewhat disagreed to the statement that "the warnings helped me detect issues early in inference". 3 participants remained neutral on the question, and 4 somewhat agreed. The participant who disagreed with this, however, pointed out that they "[...] only really look[ed] at the warnings after inference".

In contrast, another participant stated that, "[...] their dynamicity really reflects the changing nature of the sampling process and how it should be interpreted". 

We take from these results that the overall perception of warnings during inference is positive, but that their usefulness during inference is less clear than after inference. This could potentially be improved upon in future framework instantiations through better heuristics for raising warnings during inference.

\subsubsection{Some users felt misguided or were missing information}
Fig.~\ref{fig:warnings} shows that most participants disagreed with the statement that they "[...] felt misguided by some warnings" and only one participant somewhat agreed to this. Nevertheless when asked whether they felt misguided by warnings during or after inference, specifically (multiple choice), 4 participants mentioned that they felt misguided by warnings during inference, 1 after inference and 14 that they never felt misguided. This suggests that the warnings system might need adjustments in some areas. 

This is also reflected by statements made by two participants. One participant expressed that "in task three it told [them] to reformulate but it was missing what this would change" and another mentioned that "[...] it is somewhat counter intuitive, that e.g. the same warning appears 8 times, if the variable is of dim=8 [...]". Both statements express potential improvements we agree with. More detailed warnings could further improve usability and interpret-ability and reducing the amounts of almost duplicate warnings displayed could reduce clutter and improve clarity.

\section{Related Work}

\paragraph{Debugging Probabilistic Programs:}
Gorinova et al. proposed a live, multiple-representation editor for probabilistic programs targeting novices. The editor features a code editor, a model graph and prior as well as posterior distribution charts \cite{Gorinova2016ANovices}. They further showed effectiveness of multiple representations for probabilistic program comprehension in their study. Our work draws from their work and also features multiple representations. While in their work, \textit{live} refers to dynamically updating the representations when source code changes, our work features live updates in the context of inference. Furthermore, our work enables deep insights into the inference process and offers visual debugging aids as well as contextualized warnings and information on resolving potential issues that is model, inference algorithm and PPL specific. 
Hoppen et al.\cite{HoppenAPrograms} proposed a debugger for probabilistic programs that, in contrast to our work, resembles classical step-by-step debugging workflows more closely. It allows setting breakpoints in a probabilistic program and shows current values as well as the distribution from which the value was drawn. Their work aims to improve comprehension of probabilistic code execution and debugging classical logic errors stemming from e.g. random numbers of loop iterations.  
\textit{DEPP} \cite{Nandi2017DebuggingPrograms} is another debugger for probabilistic programs that is closer to our work which features a combination of static and dynamic analysis. In contrast to our work, their work does not describe a \textit{PPL} independent approach, but relies on a custom \textit{PPL} implemented by them. Furthermore, they do not provide online inference insights in their described systems, which is a key innovation in our approach. 
SixthSense \cite{Dutta2022SixthSense:Learning} is another work targeting probabilistic program debugging. In contrast to our work, SixthSense aims to predict model convergence within a specified number of steps ahead-of-run. Similar to our work they offer warnings for likely causes. In contrast to our work, they aim at prediction. Our work on the other hand gives deep insights into the real inference process by providing information and analysis tools at- and after-run-time. 
ShinyStan \cite{Gabry2022ShinyStan} is an interactive diagnostic tool with a graphical user interface for analyzing Bayesian posteriors after inference. It features interactive plots and tables commonly used for identifying MCMC inference problems. In contrast to our work it does not feature online analysis of Bayesian inference and does not incorporate details from the model or inference algorithm. Our framework further suggests potential problems based on these informations to reduce the debugging exploration space.

\paragraph{Bug Patterns in Probabilistic Programs and MCMC Diagnostic Work:}
Lieberman et al.~\cite{LiebermanDebuggingResearch} explain the necessity for a dedicated debugging framework for probabilistic programming. Hamada et al.~\cite{Hamada2022BugSystems} explores bug patterns in PyMC programs and extracts four common bug categories. 
Donovan et al.~\cite{Donovan2019MCMCApproaches} gives general guidelines with clear questions and answers targeted to novices for diagnosing MCMC inference. Zitzmann et al. identifies the effective sample size as a possible stopping criterion for MCMC algorithms \cite{Zitzmann2021UsingMplus}. Vats et al.~\cite{Vats2020AnalyzingOutput} presents important MCMC diagnostics, many of which are used in our approach, and an exemplary workflow incorporating these diagnostics.
Gabry et al. show a visualization based debugging workflow \cite{GabryVisualizationWorkflow}, which was inspirational for the visualization selection in our work. \cite{Gorinova2020AutomaticPrograms} showed how funnels in probabilistic models can be identified and proposed an automatic reparameterisation. Before that, it was shown that there is no general optimal parameterisation and centered, as well as non-centered, have both potential case dependent benefits and drawbacks~\cite{Papaspiliopoulos2007AModels}.

\paragraph{Other Work in Online Diagnostics and Visualizations:} Tensorboard~\cite{MartinAbadi2015Systems} is a well known tool for online diagnostics and visualizations in deep learning  and was an inspiration to the live inference analysis presented in this paper. Glinda~\cite{Glinda} is a complete system for supporting data science that features a domain specific language, live programming and a GUI. They showed the benefits of interactive plots and live updates in the domain of data science.  

\section{Conclusion}
In this work, we propose a novel approach to debugging \textit{probabilistic programs}. To the best of our knowledge, we are the first to realize online analysis of the Bayesian inference process. We further describe a holistic debugging approach combining model visualizations, statistics, and analysis plots with warnings and directions. We implemented this approach in \toolname{} for MCMC methods and evaluated its effectiveness in a user study with 18 participants. The study showed that our approach marks a significant advancement for debugging probabilistic models that future work can build upon.

\begin{acks}
    The authors gratefully acknowledge the financial support under the scope of the COMET program within the K2 Center “Integrated Computational Material, Process and Product Engineering (IC-MPPE)” (Project No 886385). This program is supported by the Austrian Federal Ministries for Economy, Energy and Tourism (BMWET) and for Innovation, Mobility and Infrastructure (BMIMI), represented by the Austrian Research Promotion Agency (FFG), and the federal states of Styria, Upper Austria and Tyrol. 
    
    Additionally this research has been partially funded by the European Union through the CLOUDSTARS project (101086248).
\end{acks}

\bibliographystyle{IEEEtran}
\bibliography{references_final.bib}

\end{document}